# Measuring Mountains on the Moon

Sunil K. Chebolu

Ancient Greek astronomers performed several important astronomical calculations. They estimated the circumference of the earth, the distance to the moon, the distance to the sun, the diameters of the sun and moon, and more. It is noteworthy that all these calculations were based on naked-eye observations (made during the summer solstice and solar and lunar eclipses) and basic facts in Euclidean geometry.

The problem of measuring heights of mountains on the moon, however, did not arise in Greek astronomy because of the prevailing Aristotelian belief that all heavenly bodies, including the moon, were perfect spheres. This belief was overturned in 1609, the moment Galileo first used his telescope to view the heavens. Much to his surprise, he found mountains, craters, and valleys on the lunar surface. He also observed the moons of Jupiter and phases of Venus. These observations, which could not be explained in the geocentric (Earth-centered) model made the heliocentric (sun-centered) model so compelling that even the strong opponents of the heliocentric model had to accept it, because seeing is believing. This transition marked the end of an era in astronomy.

On the night of May 22, 2018, in Normal, Illinois, I enjoyed viewing the moon in my backyard, much as Galileo did four centuries earlier. The moon was clear and beautiful in her first-quarter phase, and the viewing conditions were perfect to examine the topography of the lunar surface. I saw craters, mountains, and shadows of these objects on the moon. It was a wonderful visual treat.

Figure 1 shows a photograph of the moon at approximately 10:00 p.m. taken through my Celestron NexStar 130 SLT with a 5-mm eyepiece and a 13% Moon filter.

After the observation, I used Galileo's techniques to compute the height of one of these mountains on the moon, the details of which I now share.

## Galileo's Lunar Geometry

Galileo's idea is simple but ingenious. It is based on the observation that the rays of the sun, at dawn, hit the tips of the high peaks on the moon before they reach the terrain below them. This is due to the spherical nature of the moon, which spins on its axis. It also explains why we see some tiny and isolated luminous parts in the dark side of the moon close to the *terminator* (the imaginary arc dividing the bright and dark sides of the moon) such as the bright spots labeled A and B in figure 1. Moreover, the rays of the sun are tangent to the surface of the moon at the terminator. This geometry is illustrated (and clearly not drawn to scale) in figure 2.

**Figure 1.** Two mountain peaks (labeled A and B) on the surface of the moon.

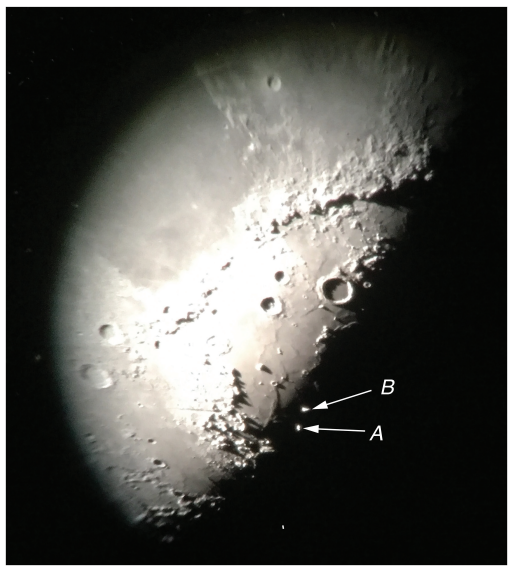

**Figure 2.** The geometry of the sun's rays striking a mountain on the moon.

**Figure 3.** Earth, the mountain on the moon, and the nearest point on the terminator form a right triangle.

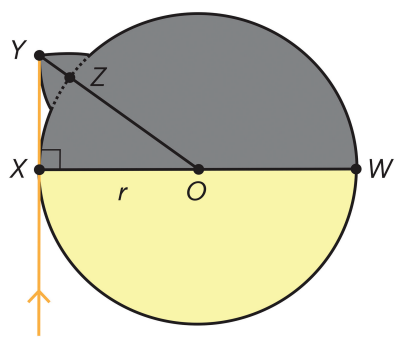
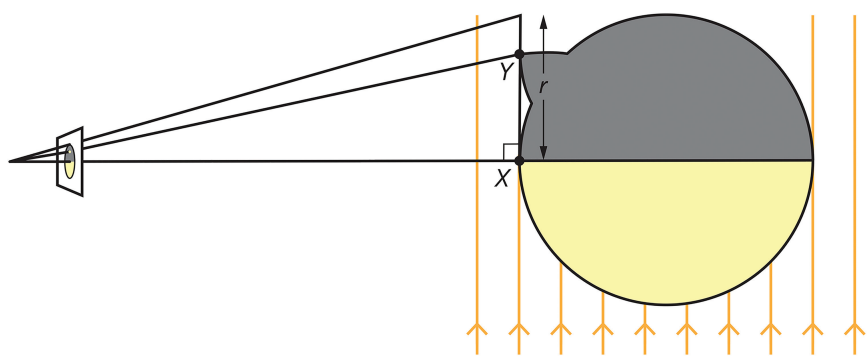

In the figure, the lunar disc is represented by a circle with center $O$. The line $XY$ is light ray from the sun tangent to the circle at $X$ and passing through the mountain peak $Y$. Diameter $XW$ is the terminator (on first quarter of the moon). The point $Z$ is on $OY$ so that $|OX|=|OZ|=r$ is the radius of the moon. The radial segment $OX$ is perpendicular to the tangent $XY$—a well-known fact from geometry. So, applying the Pythagorean theorem to the right triangle $OXY$ gives the following equations for the height $H$ of the mountain $YZ$:

$$\begin{aligned}H &= |OY|-|OZ| \\ &= \sqrt{|OX|^2+|XY|^2}-|OZ| \\ &= |OX|\sqrt{1+\frac{|XY|^2}{|OX|^2}}-|OZ| \\ &= r\sqrt{1+\left(\frac{|XY|}{r}\right)^2}-r \\ &= r\sqrt{1+\left(\frac{1}{d}\right)^2}-r,\end{aligned}$$

where $d = r/|XY|$. It is well known (since antiquity) that $r \approx 1{,}079$ miles. So, to find $H$, it is enough to find the value of $d$.

Computing $d$ directly seems hard, but fortunately it can be computed indirectly from the photograph because of the similarity of figures! Note that on the first-quarter phase, earth, the mountain on moon, and the nearest point on the terminator form a right triangle (see figure 3). Therefore, distances on the moon will be scaled down proportionately on the photograph. In particular, the value $d = r/|XY|$ is the same when computed using measurements on the moon as when using measurements on the photograph. Measuring the former is not possible, but measuring the latter is easy.

For the peak A in figure 1, the values on the original photograph are 30 mm and 2 mm, respectively. This yields $d = 15$. Substituting $d = 15$ and $r = 1{,}079$ into the formula for $H$ gives the height of peak A as approximately 2.39 miles.

Although this is only a rough estimate, this method of Galileo's is correct and is a great example to illustrate the power of mathematics. A simple astronomical observation—the presence of tiny luminous spots in the dark side of the moon—when viewed in the right mathematical framework answered a problem that seemed impossible to solve: What is the height of a mountain on the moon? ●

*Sunil K. Chebolu is a professor of mathematics at Illinois State University. When his mathematical battery is dying, he likes to charge it by either playing his guitar or exploring the night sky with his telescope.*